\documentclass[preprintnumbers,amsmath,amssymb,floatfix,14pt,prd,twocolumn,showpacs,
superscriptaddress,nofootinbib]{revtex4}
\usepackage{graphicx}
\usepackage{epsfig}
\usepackage{bm}
\usepackage{amsfonts}

\begin{document}

\title{Attractor Solutions in  $f(T)$ Cosmology}

\author{\textbf{ Mubasher Jamil}} \email{mjamil@camp.nust.edu.pk}
\affiliation{Center for Advanced Mathematics and Physics (CAMP),\\
National University of Sciences and Technology (NUST), H-12,
Islamabad, Pakistan}
\affiliation{Eurasian International Center
for Theoretical Physics, Eurasian National University, Astana
010008, Kazakhstan}
\author{\textbf{ D. Momeni}}
\email{d.momeni@yahoo.com; davidmathphys@yahoo.co.uk}
 \affiliation{Eurasian International Center
for Theoretical Physics, Eurasian National University, Astana
010008, Kazakhstan}
\author{\textbf{ Ratbay Myrzakulov}}
\email{rmyrzakulov@gmail.com;
rmyrzakulov@csufresno.edu}\affiliation{Eurasian International Center
for Theoretical Physics, Eurasian National University, Astana
010008, Kazakhstan}

\begin{abstract}
{\bf Abstract:}
 In this paper, we explore the cosmological
implications of interacting dark energy model in a torsion based
gravity namely $f(T)$. Assuming dark energy interacts with dark
matter and radiation components, we examine the stability of this
model by choosing different forms of interaction terms. We consider
three different forms of dark energy: cosmological constant,
quintessence and phantom energy. We then obtain several attractor
solutions for each dark energy model interacting with other
components. This model successfully explains the coincidence problem
via the interacting dark energy scenario.

\end{abstract}

\pacs{04.20.Fy; 04.50.+h; 98.80.-k} \maketitle

\newpage
\section{Introduction}

Its now well-accepted in astrophysics community that the observable
universe is in a phase of rapid expansion whose rate of expansion is
increasing, so called `accelerated expansion'. This conclusion has
been drawn by numerous recent cosmological and astrophysical data
findings of supernovae SNe Ia {\cite{c1}}, cosmic microwave
background radiations via WMAP {\cite{c2}}, galaxy redshift surveys
via SDSS {\cite{c3}} and galactic X-ray {\cite{c4}}. This phenomenon
is commonly termed `dark energy' (DE) in the literature, and
suggests that a cosmic dark fluid possessing negative pressure and
positive energy density. Although the phenomenon of dark energy in
cosmic history is very recent $z\sim0.7$, it has opened new areas in
cosmology research. Two important problems like `fine tuning' and
`cosmic coincidence' are related to dark energy. It is thought that
the most elegant solution to DE paradigm is the Einstein's
cosmological constant \cite{c7} but it cannot resolve the two
problems mentioned above. Hence cosmologists looked for other
theoretical models by considering the dynamic nature of dark energy
like quintessence scalar field \cite{quint}, a phantom energy field
\cite{phant} and f-essence \cite{f}. Another interesting set of
proposals to DE puzzle is the `modified gravity'  which was proposed
after the failure of general relativity (GR). This new set of
gravity theories passes several solar system and astrophysical tests
successfully \cite{sergei2}.

For the past few years, models based on DE interacting with dark
matter or any other exotic component have gained great impetus. Such
interacting DE models can successfully explain numerous cosmological
puzzles including dynamic DE, phantom crossing, cosmic coincidence
and cosmic age \cite{jamil} and also in good compatibility with the
astrophysical observations of cosmic microwave background, supernova
type Ia, baryonic acoustic oscillations and galaxy redshift surveys
\cite{obs}. There are some criticisms on interacting DE models for
not being favored from observations and that the usual $\Lambda$CDM
model is favorable \cite{cr}. However the thermal properties of this
model in various gravities have been discussed in literature
\cite{jamil2}. The model in which dark energy interacts with two
different fluids has been investigated in literature. In
\cite{cruz}, the two fluids were dark matter and another was
unspecified. However, in another investigation \cite{jamil78}, the
third component was taken as radiation to address the
cosmic-triple-coincidence problem and study the generalized second
law of thermodynamics. In a recent investigation, the authors
investigated the coincidence problem in loop quantum gravity with
triple interacting fluids including DE, dark matter and unparticle
\cite{jamil8}.
%%%%%%%%%%%%%%%%%%%%%%%%%%%%%%%%%%%%%%%%%%%%%%%%%%%%%%%%%%%%%%%%%

There is no need to construct a gravitational theory on a Riemannian
manifold. A manifold can be divided into two separate but connected
parts; one with Riemannian structure with a definite metric and
another part, with a non-Riemannian structure and with torsion or
non-metricity. That part which has the zero Riemannian tensor but
has non zero torsion is based on a tetrad basis, and defines a
Weitzenbock spacetime. $f(T)$ gravity is an alternative theory for
GR, defined on the Weitzenbock non-Riemannian manifold, working only
with torsion. This model firstly proposed by Einstein for unifying
the electromagnetism and the gravity. If $f(T)=T$, this theory is
called teleparallel gravity \cite{hayashi,hehl}. It has been shown
that with linear $T$, this model has many common features like GR
and is in good agreement with some standard tests of the GR in solar
system \cite{hayashi}. But introducing a general model $f(T)$ backs
to few years \cite{f(T)}. This model has many features, for example
Birkhoff's theorem has been studied in this gravity \cite{birkhoff}.
Earlier the authors in \cite{zheng} investigated perturbation in
$f(T)$ and found that the perturbation in $f(T)$ gravity grows
slower than that in Einstein general relativity. Bamba et al
\cite{bamba} studied the evolution of equation of state parameter
and phantom crossing in $f(T)$ model. Emergent universes in
chameleon $f(T)$ model is investigated in \cite{ujjal}. As a
thermodynamical view, there are many strange features. It has been
shown in $f(T)$ gravity, the famous formula of entropy-area of the
black hole thermodynamics is not valid \cite{miaoli}. The reason
goes back to the violation of the local lorentz invariance of this
theory \cite{sotirio,miaoli}. It shows that it is not possible to
use the Wald conjecture \cite{wald} for calculating the entropy as a
Noether charge in $f(T)$. The Hamiltonian formulation of $f(T)$
gravity has been studied in \cite{miao} and shown that there are
five degrees of freedom. In this paper we study the triple
coincidence problem: why we happen to live during this special epoch
when $ \rho_\Lambda \sim \rho_m \sim \rho_r$ ? \cite{nima}. We
investigate this problem in the framework of $f(T)$ gravity.

%%%%%%%%%%%%%%%%%%%%%%%%%%%%%%%%%%%%%%%%%%%%%%%%%%%%%%%%%%%%%%%%%%%%%%%%%%%%%
We follow the plan: In section II we introduce the basic equations
as an autonomous dynamical system. In section III we choose a model
for $f(T)$ gravity. In section IV we working on numerical analysis
of the stability and the evolution of the functions of the model in
details. We conclude and summarize in section V.

\section{Basic equations}
One suitable form of  action for $f(T)$ gravity in  Weitzenbock
 spacetime is  \cite{f(T)}
\begin{eqnarray}\nonumber
S=\frac{1}{2\kappa^2}\int d^4x \sqrt{e}(T+f(T)+L_m)
\end{eqnarray}
Here $e=det(e^{i}_{\mu})$, $\kappa^2=8\pi G$ and $e^{i}_{\mu}$ is the tetrad (vierbein) basis. The dynamical quantity of the model is the scalar torsion $T$ and $L_m$ is the matter Lagrangian.
We start with the Friedmann equation in this form of the $f(T)$ model \cite{f(T)}
\begin{equation}\label{1}
H^2=\frac{1}{1+2f_T}\Big( \frac{\kappa^2}{3}\rho-\frac{f}{6} \Big),
\end{equation}
where $\rho=\rho_m+\rho_d+\rho_r$, while $\rho_m$, $\rho_d$ and
$\rho_r$ represent the energy densities of matter, dark energy and
the radiation respectively.

 Another FRW equation is
\begin{equation}\label{2}
\dot H=-\frac{\kappa^2}{2}\Big(\frac{\rho+p}{1+f_T+2Tf_{TT}}\Big).
\end{equation}
For a spatially flat universe ($k=0$),  the total energy
conservation equation is
\begin{equation}\label{1a}
\dot \rho+3H(\rho+p)=0,
\end{equation}
where $H$ is the Hubble parameter, $\rho$ is the total energy
density and $p$ is the total pressure of the background fluid.

We assume a three component fluid containing matter, dark energy and
radiation having an interaction. The corresponding continuity
equations are \cite{jamil8}
\begin{eqnarray}\label{2a}
\dot \rho_d+3H(\rho_d+p_d)&=&\Gamma_1,\nonumber\\
\dot\rho_m+3H\rho_m&=&\Gamma_2,\\ \dot
\rho_r+3H(\rho_r+p_r)&=&\Gamma_3,\nonumber
\end{eqnarray}
which satisfy collectively (\ref{1a}) such that
$\Gamma_1+\Gamma_2+\Gamma_3=0$.

We define dimensionless density parameters via
\begin{equation}\label{3}
x\equiv\frac{\kappa^2\rho_d}{3H^2},\ \
y\equiv\frac{\kappa^2\rho_m}{3H^2},\ \
z\equiv\frac{\kappa^2\rho_r}{3H^2}.
\end{equation}
The continuity equations (\ref{2a}) in dimensionless variables
reduce to
\begin{eqnarray}\label{sys}
\frac{dx}{dN}&=&3x\Big(\frac{x+y+z+w_d x+w_r z}{1+f_T+2T
f_{TT}}\Big)\\&&\nonumber-3x(1+w_d)+\frac{\kappa^2}{3H^3}\Gamma_1,\\\nonumber
\frac{dy}{dN}&=&3y\Big(\frac{x+y+z+w_d x+w_r z}{1+f_T+2T
f_{TT}}\Big)\\&&\nonumber-3y+\frac{\kappa^2}{3H^3}\Gamma_2,\\\nonumber
\frac{dz}{dN}&=&3z\Big(\frac{x+y+z+w_d x+w_r z}{1+f_T+2T
f_{TT}}\Big)\\&&\nonumber-3z(1+w_r)+\frac{\kappa^2}{3H^3}\Gamma_3\label{eq3},
\end{eqnarray}
where $N\equiv\ln a$, is called the e-folding parameter. The
coupling functions $\Gamma_i$, $i=1,2,3$ are in general functions of
the energy densities and the Hubble parameter i.e.
$\Gamma_i(H\rho_i)$. The system of equations in (\ref{sys}) is
analyzed by first equating them to zero to obtain the critical
points. Next we perturb equations up to first order about the
critical points and check their stability. Below for computation, we
shall assume $w_m=0$, $w_r=\frac{1}{3}$ and $w_d$ to be a general
non-zero but negative parameter. We are interested in stable
critical points (i.e. those points for which all eigenvalues of
Jacobian matrix are negative) as these are attractor solutions of
the dynamical system.

\section{$f(T)$  model }

To avoid analytic and computation problems, we choose a suitable
$f(T)$ expression which contains a constant, linear and a non-linear
form of torsion, specifically
\begin{equation}
f(T)=2C_1 \sqrt{-T} +\alpha T+C_2,
\end{equation}
where $\alpha$, $C_1$ and $C_2$ are arbitrary constants. The first
and the third terms (excluding the middle term) has correspondence
with the cosmological constant EoS in $f(T)$ gravity \cite{mirza}.
There are many kinds of such these models, reconstructed from
different kinds of the dark energy models. For example this form (7)
may be inspired from a model for dark energy from proposed form of
the Veneziano ghost \cite{kk}. But the linear term is needed to show
the differences between our results in $f(T)$ gravity from the
Einstein gravity. Here we choose this model to simplify our
numerical computations and for easier discussion on the difference
of our results with the same results in GR. It is the minimum model,
but our equations have been written for a general $f(T)$ action. It
is possible by repeating the numerical steps as we done in this
paper, discuss the stability of other models. Recently Capozziello et al \cite{capo123} inviestigated the cosmography of $f(T)$ cosmology by using data of BAO, Supernovae Ia and WMAP. Following their interesting results, we notice that if we choose $C_2=0$, $\alpha=\Omega_{m0}$ and $C_1=\sqrt{6}H_0(\Omega_{m0}-1)$, than we can estimate the parameters of our proposed $f(T)$ model as a function of Hubble parameter $H_0$ and the cosmographic parameters and the value of matter density parameter.

\section{Analysis of stability in phase space}

In this section, we will construct four models by choosing different
coupling forms $\Gamma_i$ and analyze the stability of the
corresponding dynamical systems about the critical points. We shall
plot the phase and evolutionary diagrams accordingly. For this reason, we must find the critical points of the (\ref{sys}), and then we linearize the system near the critical points up to first order.

\subsection{Interacting model - I}

We consider the model with the following interaction terms
\begin{equation}\label{8a}
\Gamma_1=-6bH\rho_d, \ \ \Gamma_2=\Gamma_3=3bH\rho_d,
\end{equation}
where $b$ is a coupling parameter and we assume it to be a positive
real number of order unity. Thus (\ref{8a}) says that both matter
and radiation have increase in energy density with time while dark
energy loses its energy density. Therefore it is a decay of dark
energy into matter and radiation.

Using (\ref{8a}), the system (\ref{sys}) takes the form
\begin{eqnarray}
\frac{dx}{dN}&=&-3x(1+w_d)+3x\Big(\frac{x+y+z+w_d x+w_r z}{1+\alpha}\Big)-6bx,\nonumber\\
\frac{dy}{dN}&=&-3y+3y\Big(\frac{x+y+z+w_d x+w_r z}{1+\alpha}\Big)+3bx,\label{33}\\
\frac{dz}{dN}&=&-3z(1+w_r)+3z\Big(\frac{x+y+z+w_d x+w_r
z}{1+\alpha}\Big)+3bx.\nonumber
\end{eqnarray}
The critical points for this model are obtained by equating the left
hand sides of (\ref{33}) to zero. We obtain four critical points:
\begin{itemize}
\item Point $A_1:\ (\frac{(1+\alpha)(1+w_d+2b)}{1+w_d},0,0)$,
\item Point $B_1:\ (0,0,0)$,
\item Point $C_1:\  (0,(1-b)(1+\alpha),0), $
\item Point $D_1 :\  (0,0,\frac{3}{4}(1+\alpha)(\frac{4}{3}-b))$
\end{itemize}

The eigenvalues of the Jacobian matrix for these critical points
are:
\begin{itemize}
\item Point $A_1:\lambda_1=3(1+w_d+2b), \lambda_2= 3(w_d+3b),
\lambda_3= -1+3(w_d+3b),$
\item Point $B_1:\lambda_1= 3(b-1), \lambda_2=-3(b-1),
\lambda_3=-3(1+2b+w_d)$,
\item Point $C_1:\lambda_1=-1, \lambda_2=3(1-b), \lambda_3=
-3(w_d+3b)$,
\item Point $D_1:\lambda_1=1,\lambda_2=4-3b, \lambda_3=
1-3w_d-9b$
\end{itemize}
Point $A_1$ is stable when one of these conditions is satisfied:
\begin{eqnarray}
w_d<-3,\ \ b<1/18.\\
w_d\geq-3, \ \ w_d<-10/9,\ \ b<1/18.\\
w_d\geq-\frac{10}{9},\ \ w_d<0,\ \  b<-\frac{1}{2}(1+w_d).
\end{eqnarray}
 $B_1$ is an unstable critical point since if $b>1$ than
$\lambda_1>0$ but $\lambda_2<0$. $C_1$ is stable
 if $b>1,w_d>-3b$. Similarly  $D_1$ is unstable since $\lambda_1>0$.

In figures (1-6), we plot the parameters of model-I. In figure 1, we
observe that the dimensionless density parameters evolve from their
currently observed values to vanishing densities. In figure-2, we
observe similar behavior when plotted against e-folding parameter:
the density parameters evolve from their current values to zero. In
figure-1, we deal with phantom energy, in figure-3, the quintessence
case while in figure-5, the cosmological constant case.

\begin{figure}
\centering
 \includegraphics[scale=0.4] {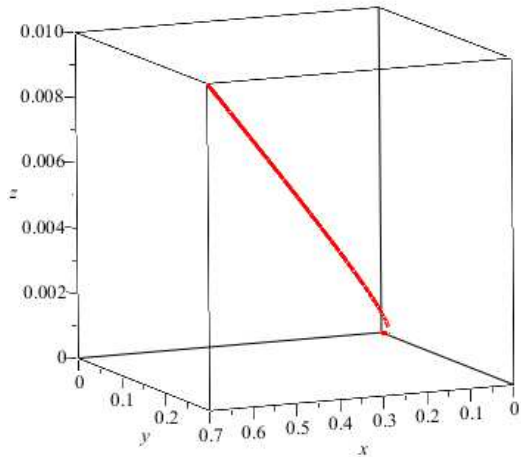}% scale goes from 0 to 1.
  \caption{ Model I: Phase space for $w_d=-1.2, b=0.5,\alpha=0.5$. It shows an attractor behavior. }
  \label{1.eps}
\end{figure}

\begin{figure}
\centering
 \includegraphics[scale=0.3] {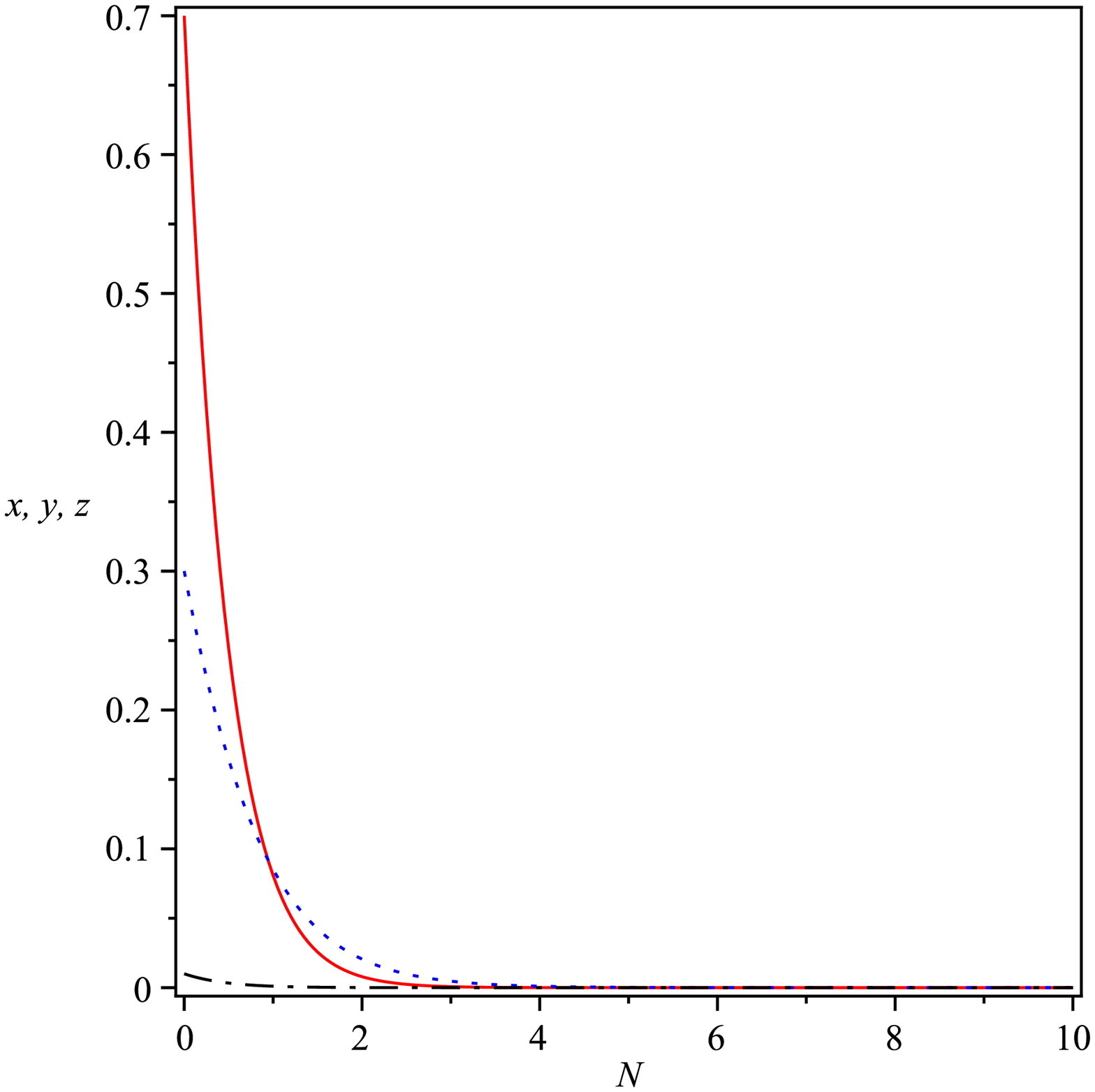}% scale goes from 0 to 1.
  \caption{ Model I: variation of $x,y,z$ as a function of the $N=\ln(a)$. The initial
  conditions chosen are $x(0)=0.7,y(0)=0.3,z(0)=0.01$,
  $w_d=-1.2$ and $b=0.5$.}
  \label{2.eps}
\end{figure}

\begin{figure}
\centering
 \includegraphics[scale=0.4] {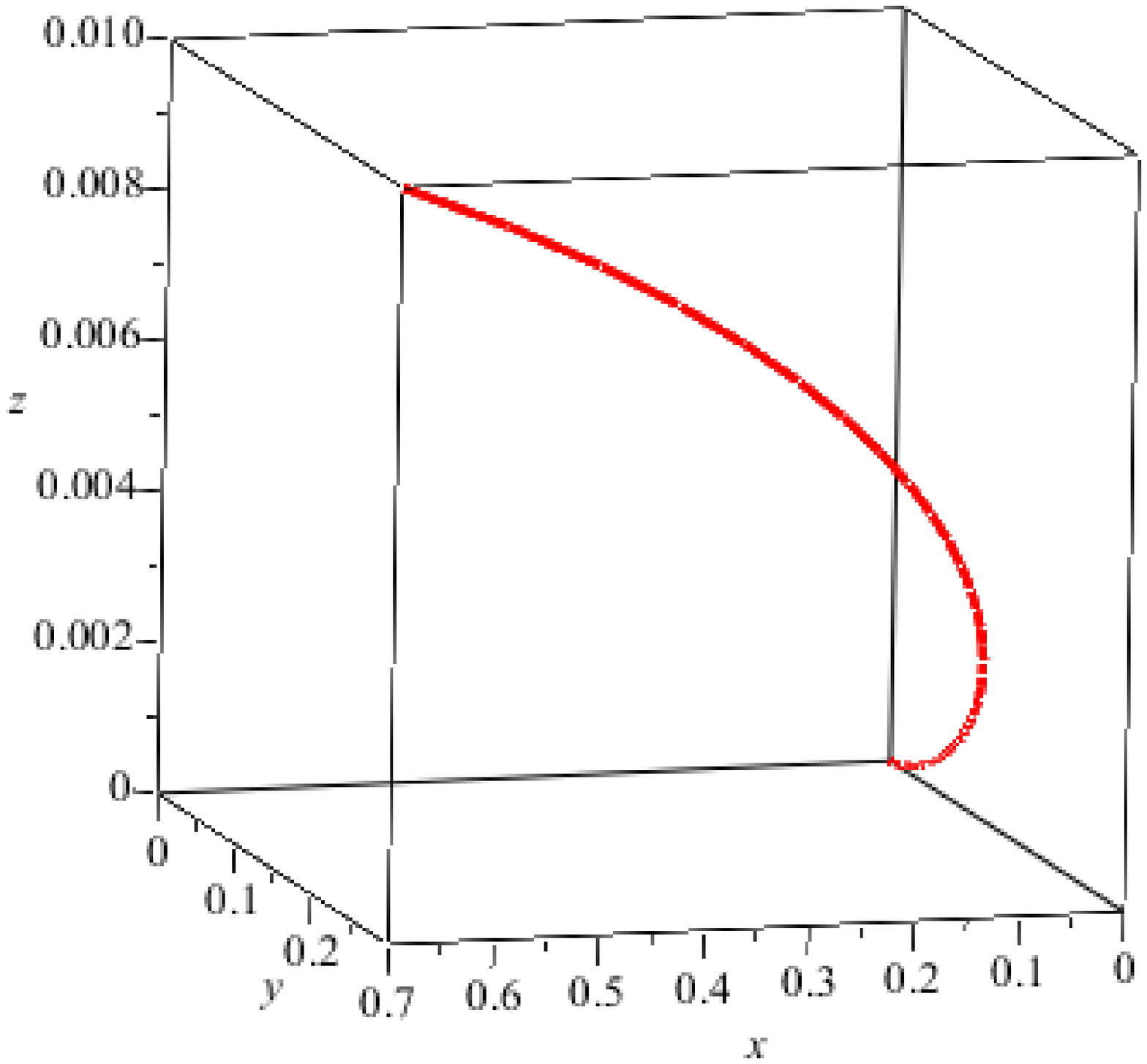}% scale goes from 0 to 1.
  \caption{ Model I: Phase space for $w_d=-0.5, b=0.5,\alpha=0.5$.}
  \label{3.eps}
\end{figure}

\begin{figure}
\centering
 \includegraphics[scale=0.3] {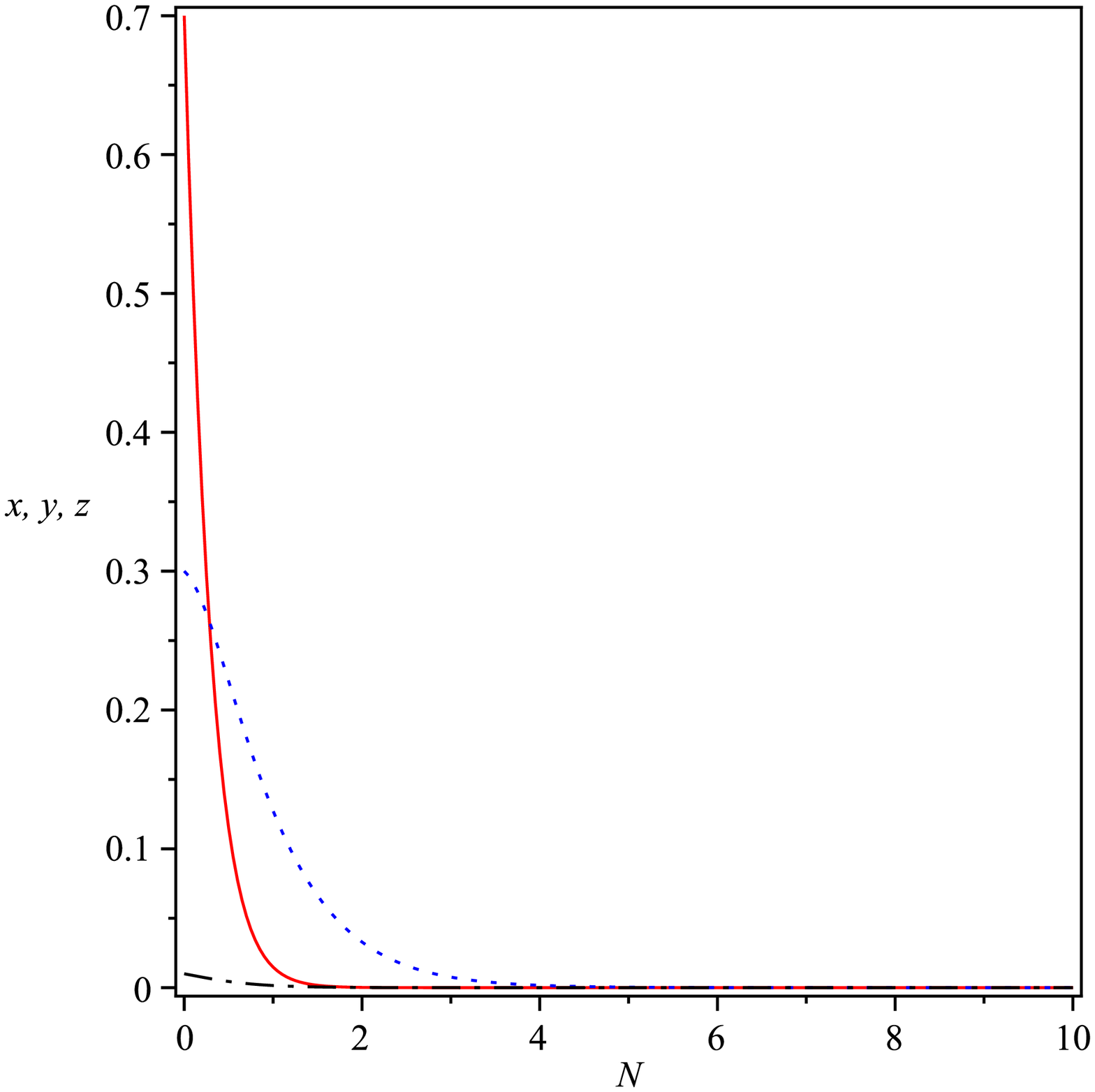}% scale goes from 0 to 1.
  \caption{ Model I: variation of $x,y,z$ as a function of the $N=\ln(a)$. The initial
  conditions chosen are $x(0)=0.7,y(0)=0.3,z(0)=0.01$,
  $w_d=-0.5$ and $b=0.5$. }
  \label{4.eps}
\end{figure}

\begin{figure}
\centering
 \includegraphics[scale=0.4] {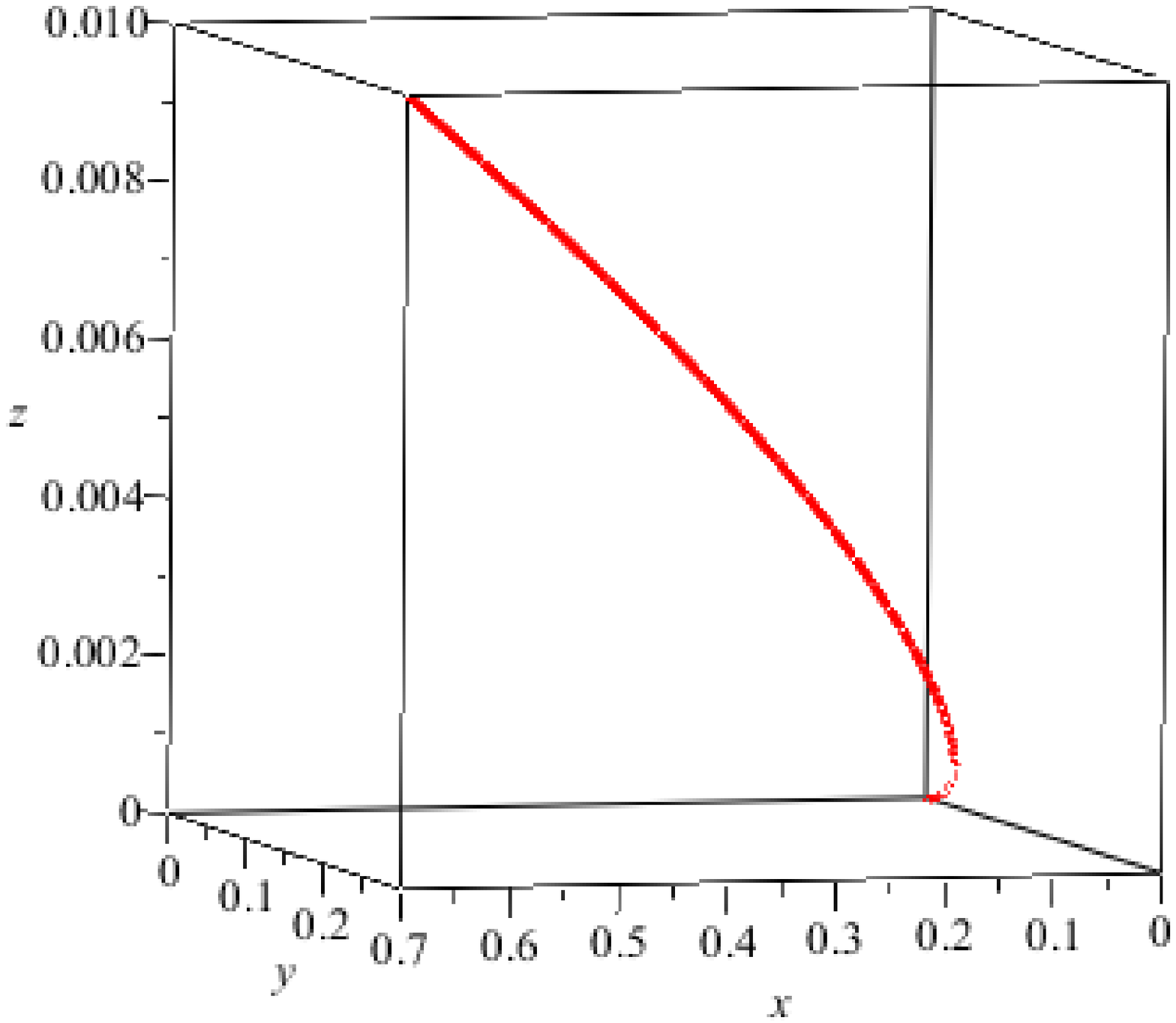}% scale goes from 0 to 1.
  \caption{ Model I:  Phase space for $w_d=-1, b=0.5,\alpha=0.5$.}
  \label{5.eps}
\end{figure}

\begin{figure}
\centering
 \includegraphics[scale=0.3] {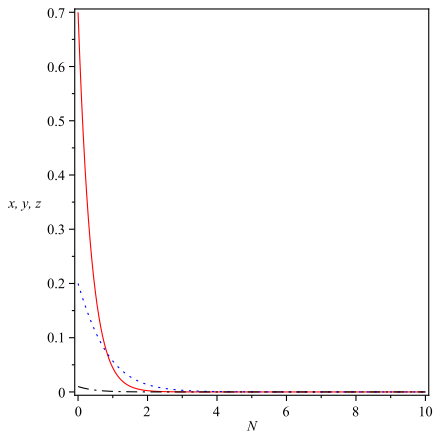}% scale goes from 0 to 1.
  \caption{Model I: variation of $x,y,z$ as a function of the $N=\ln(a)$. The initial
  conditions chosen are $x(0)=0.7,y(0)=0.3,z(0)=0.01$,
  $w_d=-1$ and $b=0.5$. }
  \label{6.eps}
\end{figure}

\subsection{Interacting model - II}

We study another model with the choice of the interaction terms
\begin{equation}\label{9a}
\Gamma_1=-3bH\rho_d,\ \ \Gamma_2=3bH(\rho_d-\rho_m),\ \
\Gamma_3=3bH\rho_m.
\end{equation}
This model effectively describes the situation when dark energy
loses energy density to matter while the radiation density increases
due to interaction with the matter.
\begin{eqnarray}\label{9}
\frac{dx}{dN}&=&3x\Big(\frac{x+y+z+w_d x+w_r z}{1+\alpha}\Big)\\&&\nonumber-3bx-3x(1+w_d),\\
\frac{dy}{dN}&=&3y\Big(\frac{x+y+z+w_d x+w_r z}{1+\alpha}\Big)\\&&\nonumber+3b(x-y)-3y,\\
\frac{dz}{dN}&=&3z\Big(\frac{x+y+z+w_d x+w_r
z}{1+\alpha}\Big)\\&&\nonumber+3by-3z(1+w_r),
\end{eqnarray}
There are four critical points:
\begin{itemize}
\item Point $A_2:\ (0,0,0)$,
\item Point $B_2:\ (0,0,1+\alpha)$,
\item Point $C_2:\ (0,(1-3b)(1+\alpha),3b(1+\alpha)$
,
\item Point $D_2:\  (-\,{\frac {3 \left( 2+w_{{d}} \right)  \left( b-w_{{d}}-1 \right)
 \left( b-w_{{d}}-7/3 \right)  \left( \alpha+1 \right) }{3\,{w_{{d}}}^
{3}+ \left( 16-3\,b \right) {w_{{d}}}^{2}+ \left( -6\,b+27 \right)
w_{ {d}}+14+{b}^{2}+b}} ,\\\nonumber-\,{\frac {b 3\left( \alpha+1
\right) \left( b-w_{{d}}-7/3 \right)
 \left( b-w_{{d}}-1 \right) }{3\,{w_{{d}}}^{3}+ \left( 16-3\,b
 \right) {w_{{d}}}^{2}+ \left( -6\,b+27 \right) w_{{d}}+14+{b}^{2}+b}}
\\,\,{\frac { 3\left( \alpha+1 \right)  \left( b-w_{{d}}-1 \right) {b}^{2
}}{14+16\,{w_{{d}}}^{2}-6\,w_{{d}}b+{b}^{2}-3\,{w_{{d}}}^{2}b+3\,{w_{{
d}}}^{3}+27\,w_{{d}}+b}}
)$
\end{itemize}
The eigenvalues of the Jacobian matrix for these critical points
are:
\begin{itemize}
\item Point $A_2:\lambda_1= -4, \lambda_2= -3(1+b),\lambda_3= 3(1+w_d-b),$
\item Point $B_2:\lambda_1= 4, \lambda_2=1-3b,\lambda_3=7+3(w_d-b)$,
\item Point $C_2:\lambda_1=3(1+b), \lambda_2=3b-1,\lambda_3= 3(w_d+2)$,
\item Point $D_2:\lambda_1=-3(w_d+2), \lambda_2=3(b-w_d-1),\lambda_3= -7-3(b-w_d)$
\end{itemize}
 $A_2$,  $D_2$ are conditionally stable if $b>1+w_d$ (for $A_2$) and  $w_d>-2$ and
$b<1+w_d$   (for $D_2$) . But $B_2$ and $C_2$ are
unstable since $\lambda_1>0$.

\begin{figure}
\centering
 \includegraphics[scale=0.4] {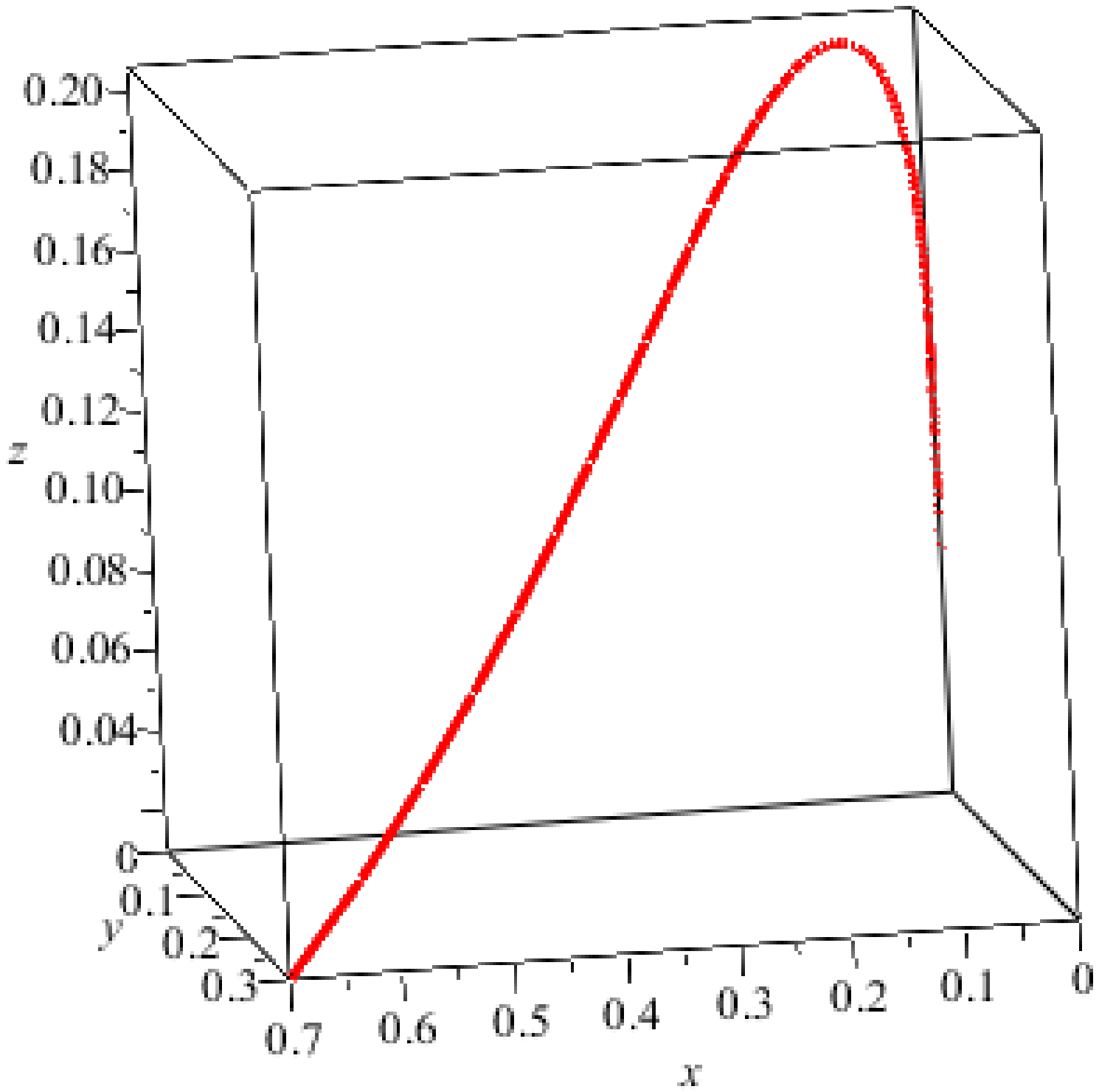}% scale goes from 0 to 1.
  \caption{ Model II: Phase space for $w_d=-1.2, b=0.5,\alpha=0.5$. }
  \label{7.eps}
\end{figure}

\begin{figure}
\centering
 \includegraphics[scale=0.3] {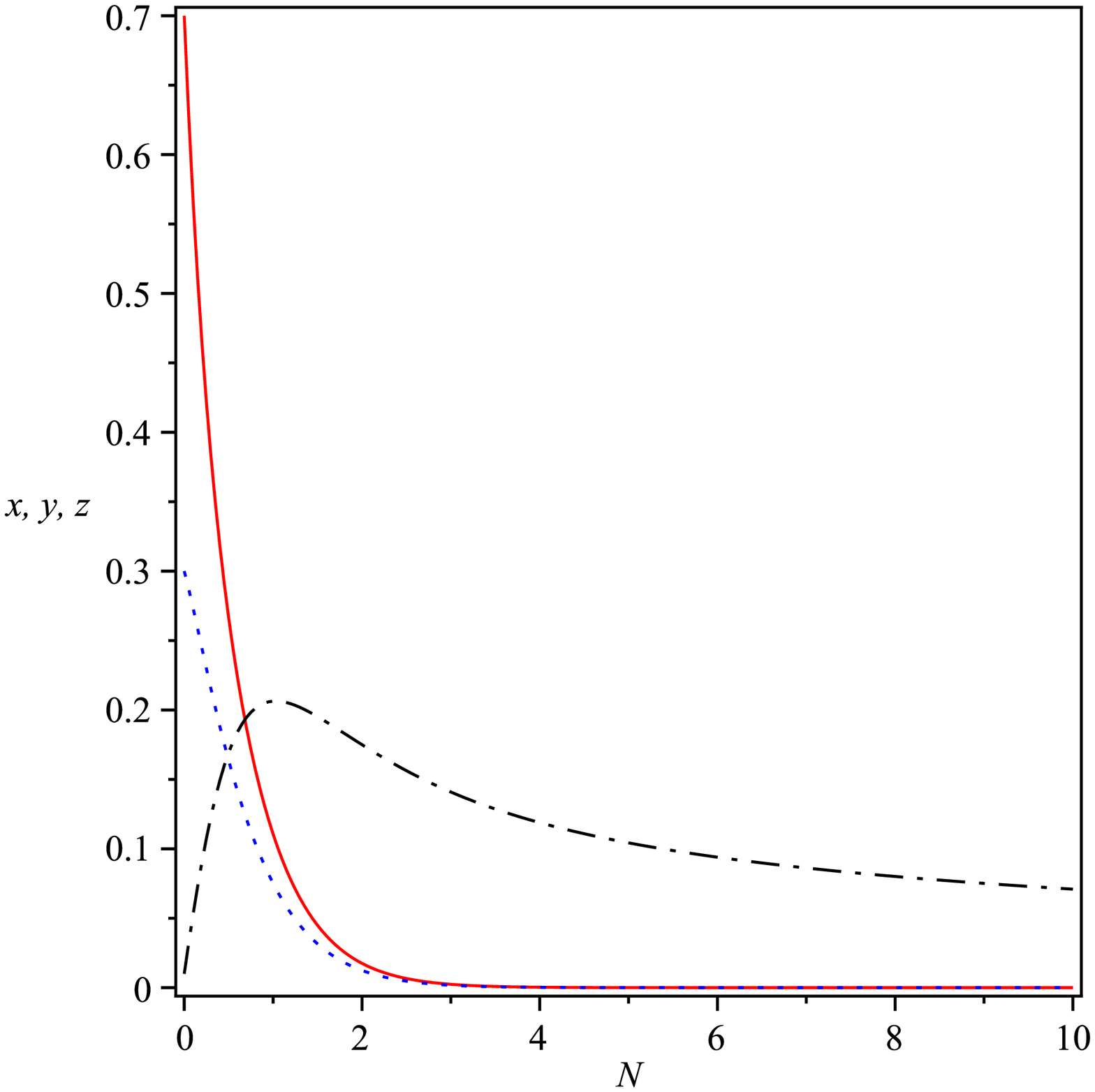}% scale goes from 0 to 1.
  \caption{ Model II: variation  of $x,y,z$ as a function of the $N=\ln(a)$. The initial
  conditions chosen are $x(0)=0.7,y(0)=0.3,z(0)=0.01$,
  $w_d=-1.2$ and $b=0.5$. }
  \label{8.eps}
\end{figure}

\begin{figure}
\centering
 \includegraphics[scale=0.4] {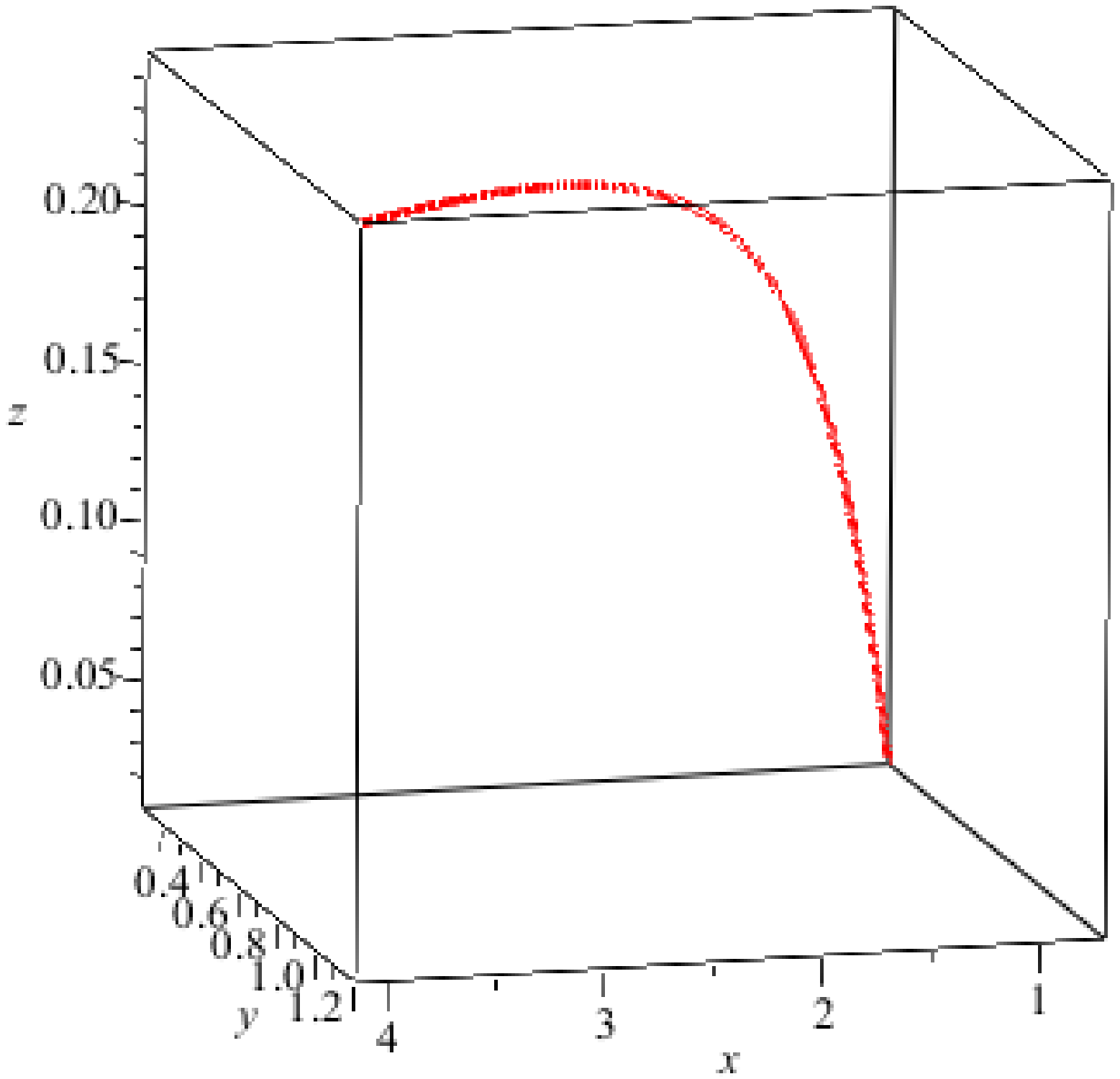}% scale goes from 0 to 1.
  \caption{ Model II: Phase space for $w_d=-0.5, b=0.5,\alpha=0.5$.}
  \label{9.eps}
\end{figure}

\begin{figure}
\centering
 \includegraphics[scale=0.3] {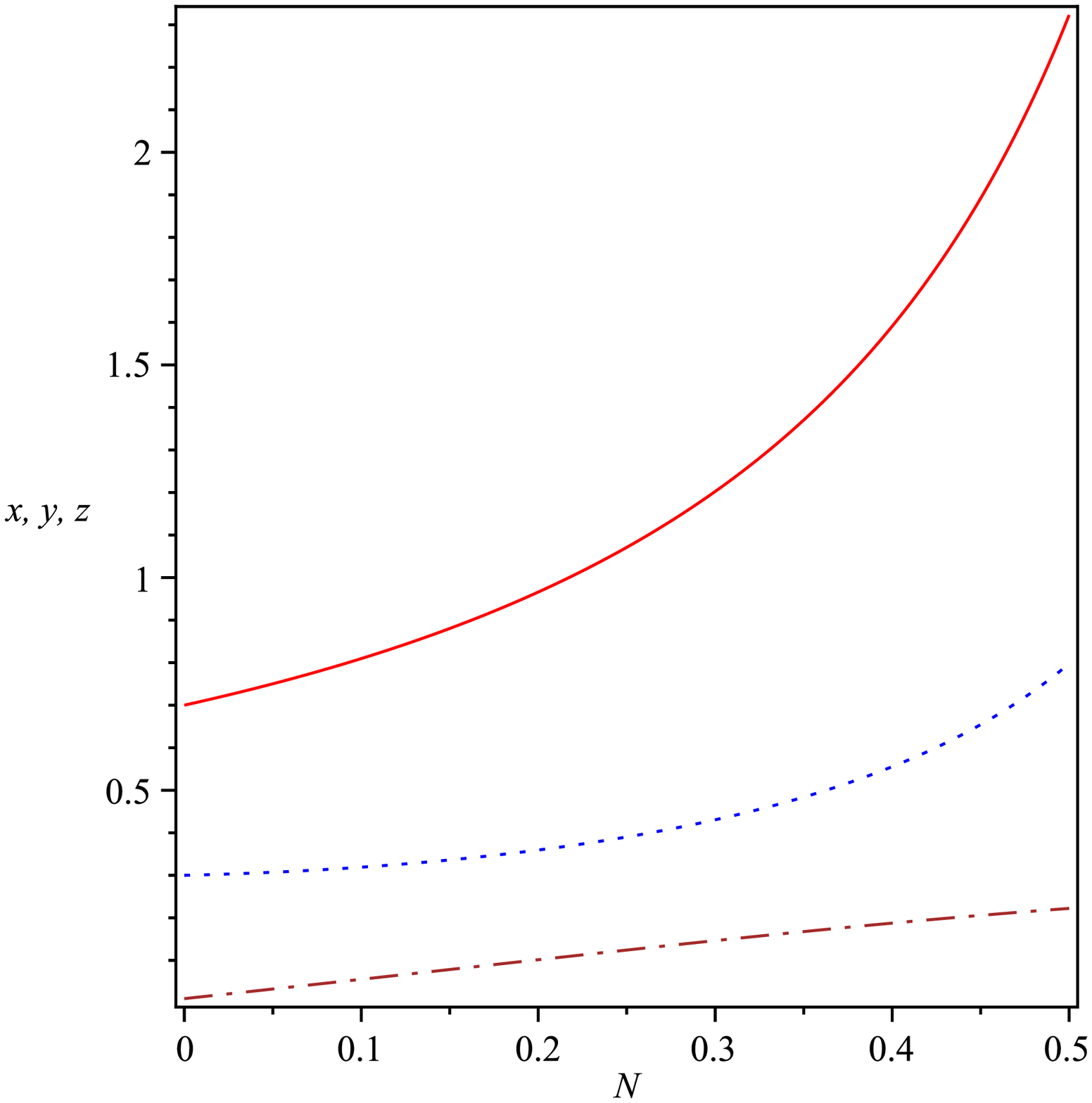}% scale goes from 0 to 1.
  \caption{Model II: variation of $x,y,z$ as a function of the $N=\ln(a)$. The initial
  conditions chosen are $x(0)=0.7,y(0)=0.3,z(0)=0.01$,
  $w_d=-0.5$ and $b=0.5$. }
  \label{10.eps}
\end{figure}

\begin{figure}
\centering
 \includegraphics[scale=0.4] {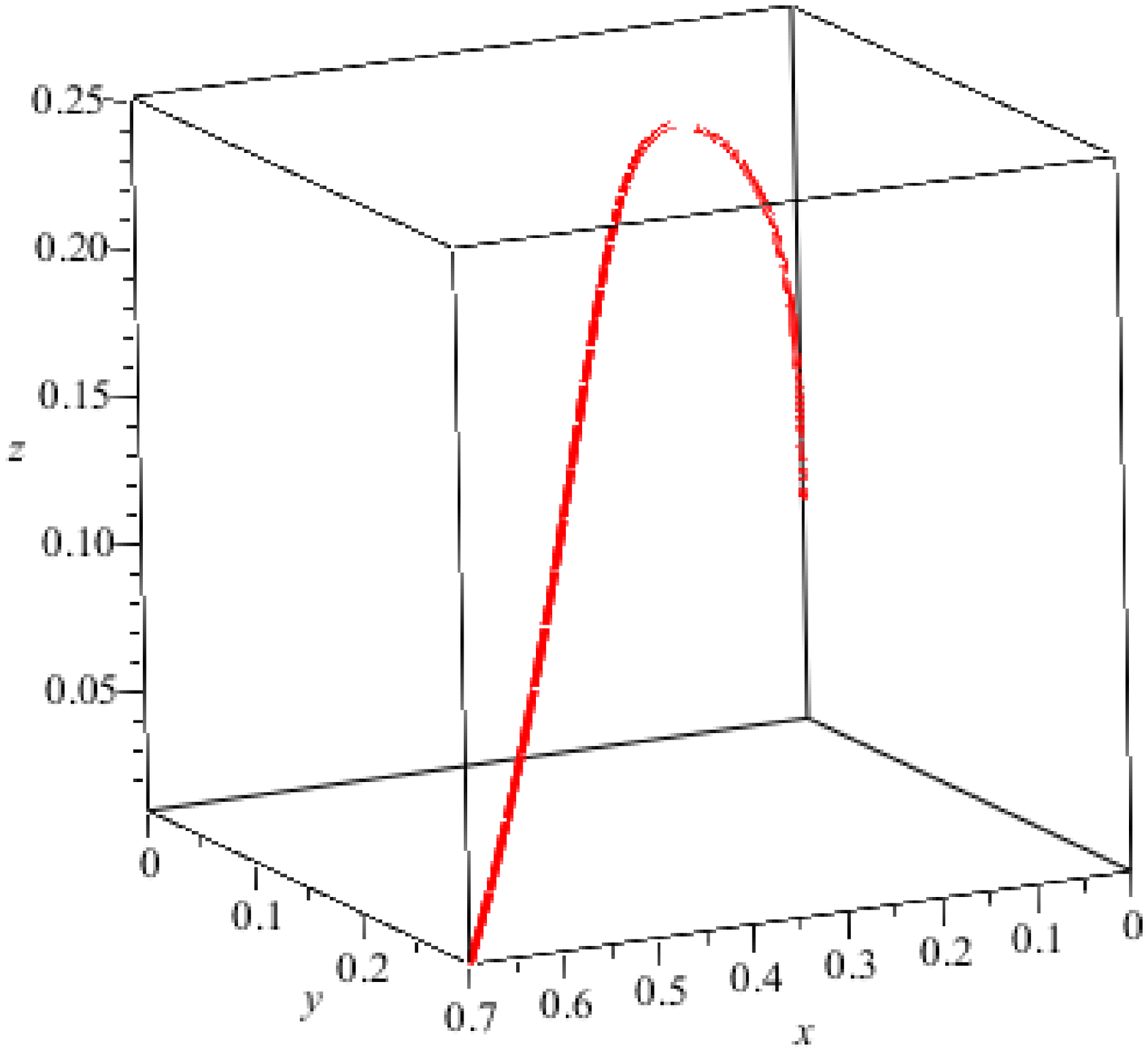}% scale goes from 0 to 1.
  \caption{Model II: Phase space for $w_d=-1, b=0.5,\alpha=0.5$.}
  \label{11.eps}
\end{figure}

\begin{figure}
\centering
 \includegraphics[scale=0.3] {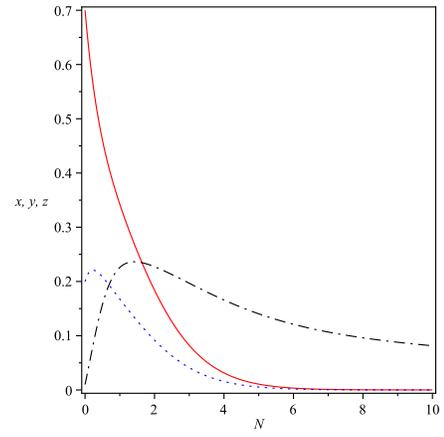}% scale goes from 0 to 1.
  \caption{ Model II: variation of $x,y,z$ as a function of the $N=\ln(a)$. The initial
  conditions chosen are $x(0)=0.7,y(0)=0.3,z(0)=0.01$,
  $w_d=-1$ and $b=0.5$.}
  \label{12.eps}
\end{figure}

In figures (7-12), we show the dynamics of Model-II. Attractor
solutions are shown in figures 7,9 and 11. In figure 8, we see that
the energy density of dark energy decays like quintessence. Also the
energy density of radiation first increases till $N\sim1.6$ and then
starts decreasing. The matter density always decreases and
approaches zero nearly $N\sim3$. In figure 10, the energy density of
dark energy increases rapidly behaving like phantom energy, energy
density of matter rises almost exponentially at later times, while
radiation density increases slower compared to both matter and dark
energy. These novel behaviors appear on account of interaction
between three components. In figure 12, the dark energy density
behaves like quintessence, while matter and radiation density falls
with expansion.

\subsection{Interacting Model - III}

 Let us take the interaction terms
\cite{jamil8}
\begin{equation}\label{12}
\Gamma_1=-6b\kappa^2 H^{-1}\rho_d\rho_r, \ \
\Gamma_2=\Gamma_3=3b\kappa^2 H^{-1}\rho_d\rho_r.
\end{equation}
The system in (\ref{sys}) takes the form
\begin{eqnarray}
\frac{dx}{dN}&=&3x\Big(\frac{x+y+z+w_dx+w_r z}{1+\alpha}\Big)
-3x-3w_dx-18bxz,\nonumber\\
\frac{dy}{dN}&=&3y\Big(\frac{x+y+z+w_dx+w_ rz}{1+\alpha}\Big)-3y+9bxz,\\
\frac{dz}{dN}&=&3z\Big(\frac{x+y+z+w_dx+w_r
z}{1+\alpha}\Big)-3z-3w_rz+9bxz.\nonumber
\end{eqnarray}
There are six critical points:
\begin{itemize}
\item Point $A_3:\ (0,0,0)$,
\item Point $B_3:\ (0,1+\alpha,0)$,
\item Point $C_3:\ (0,0,1+\alpha) $,
\item Point $D_3: \ (1+\alpha,0,0)$,
\item Point $E_3: \ (\frac{4}{9b},-\frac{2(1+w_d)}{9b},-\frac{1+w_d}{6b})$,
\item Point $F_3:  \ \Big(\,{\frac { \left( w_{{d}}+6\,b\alpha+6\,b-\frac{1}{3} \right) }{
b \left( 6\,b\alpha+6\,b-\frac{1}{3}+2\,w_{{d}} \right)
}},\,\\\nonumber{\frac {-3\,b\alpha\,-3\,b+36\,{b}^{2}\alpha+18\,{b
}^{2}{\alpha}^{2}+9\,w_{{d}}b\alpha+18\,{b}^{2}+{w_{{d}}}^{2}-2\,w_{{d
}}\frac{1}{3}+9\,w_{{d}}b+\frac{1}{9}}{3b \left(
6\,b\alpha+6\,b-\frac{1}{3}+ 2\,w_{{d}} \right) }},\\-\,{\frac
{w_{{d}} \left( 3\,b\alpha+3\,b-\frac{1}{3}+w_{{d}} \right) } {3b
\left( 6\,b\alpha+6\,b-\frac{1}{3}+2\,w_{{d}} \right) }}\Big)$
\end{itemize}
\begin{figure}
\centering
 \includegraphics[scale=0.4] {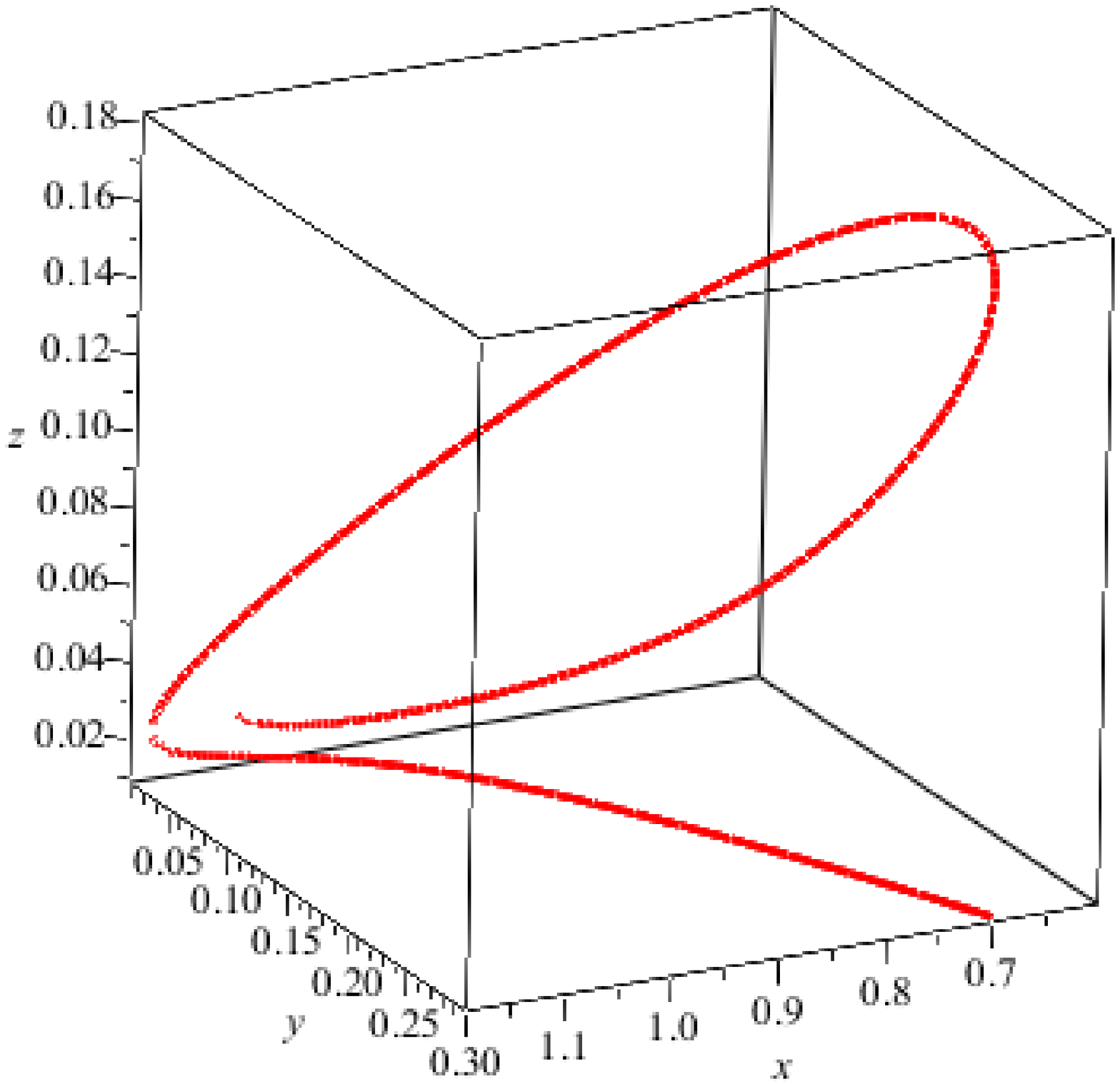}% scale goes from 0 to 1.
  \caption{ Model III: Phase space for $w_d=-1.2, b=0.5,\alpha=0.5$. }
  \label{13.eps}
\end{figure}

\begin{figure}
\centering
 \includegraphics[scale=0.3] {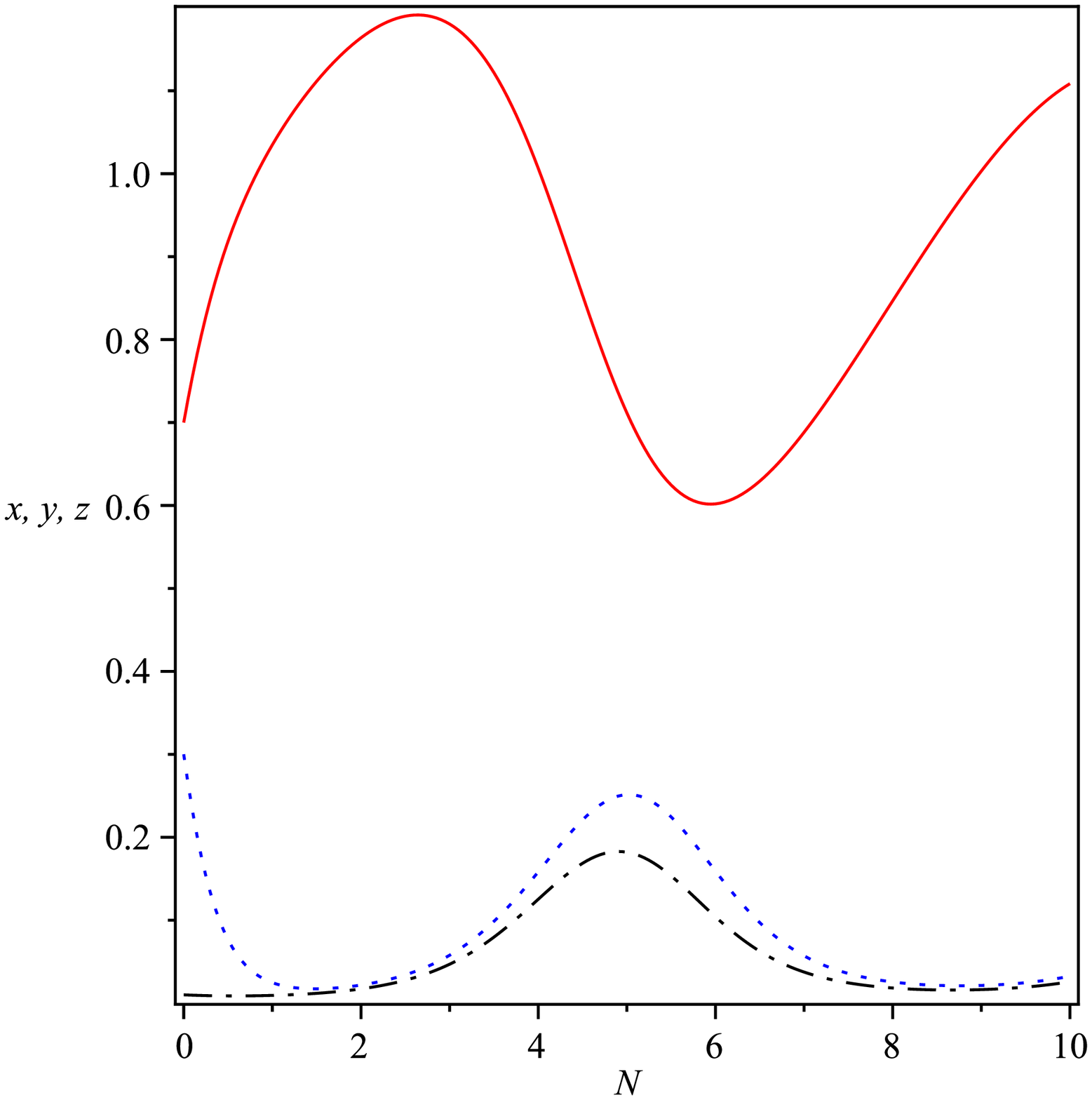}% scale goes from 0 to 1.
  \caption{ Model III: variation of $x,y,z$ as a function of the $N=\ln(a)$. The initial
  conditions chosen are $x(0)=0.7,y(0)=0.3,z(0)=0.01$,
  $w_d=-1.2$ and $b=0.5$. }
  \label{14.eps}
\end{figure}

\begin{figure}
\centering
 \includegraphics[scale=0.4] {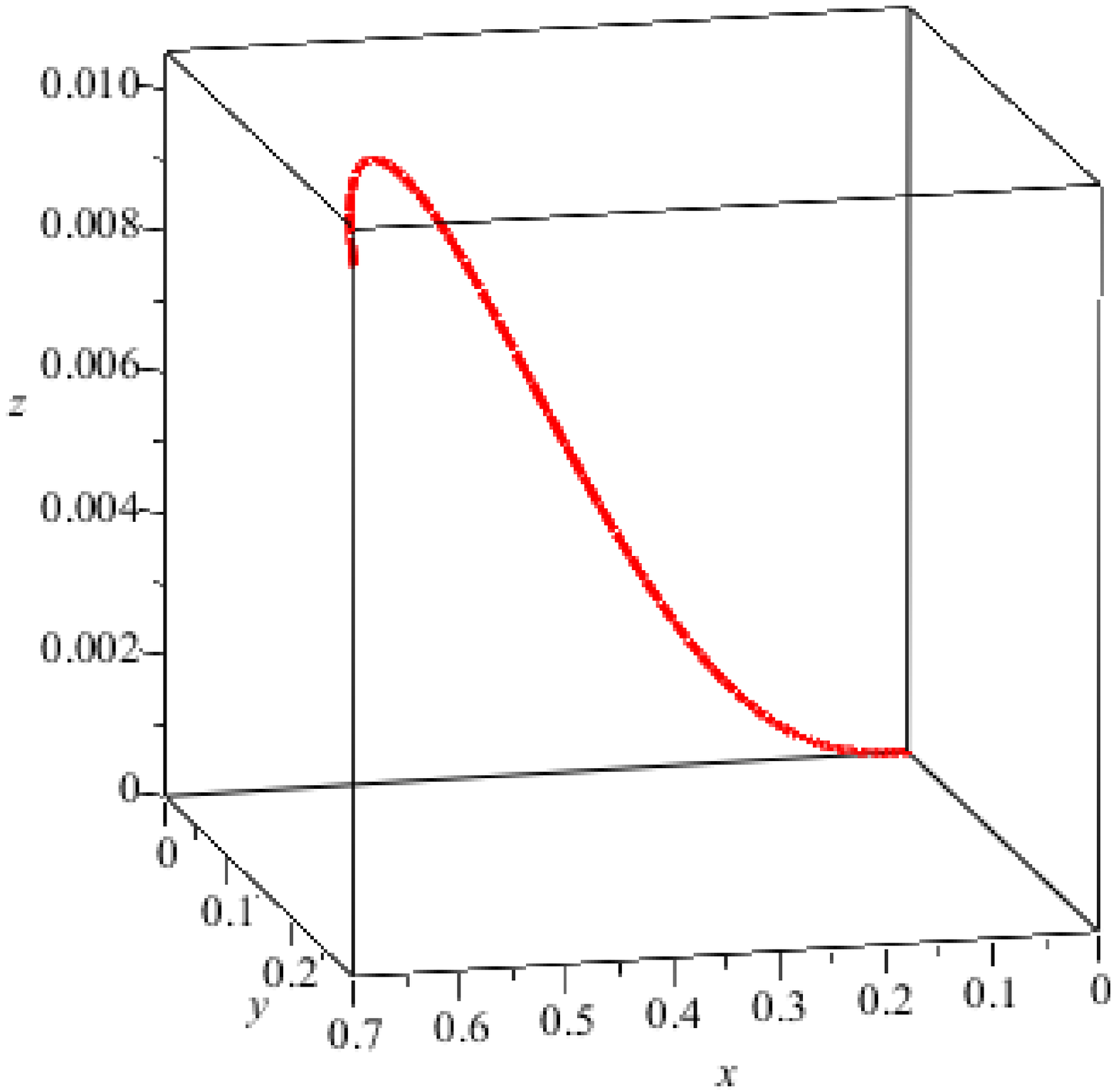}% scale goes from 0 to 1.
  \caption{Model III: Phase space for $w_d=-0.5, b=0.5,\alpha=0.5$.}
  \label{15.eps}
\end{figure}

\begin{figure}
\centering
 \includegraphics[scale=0.3] {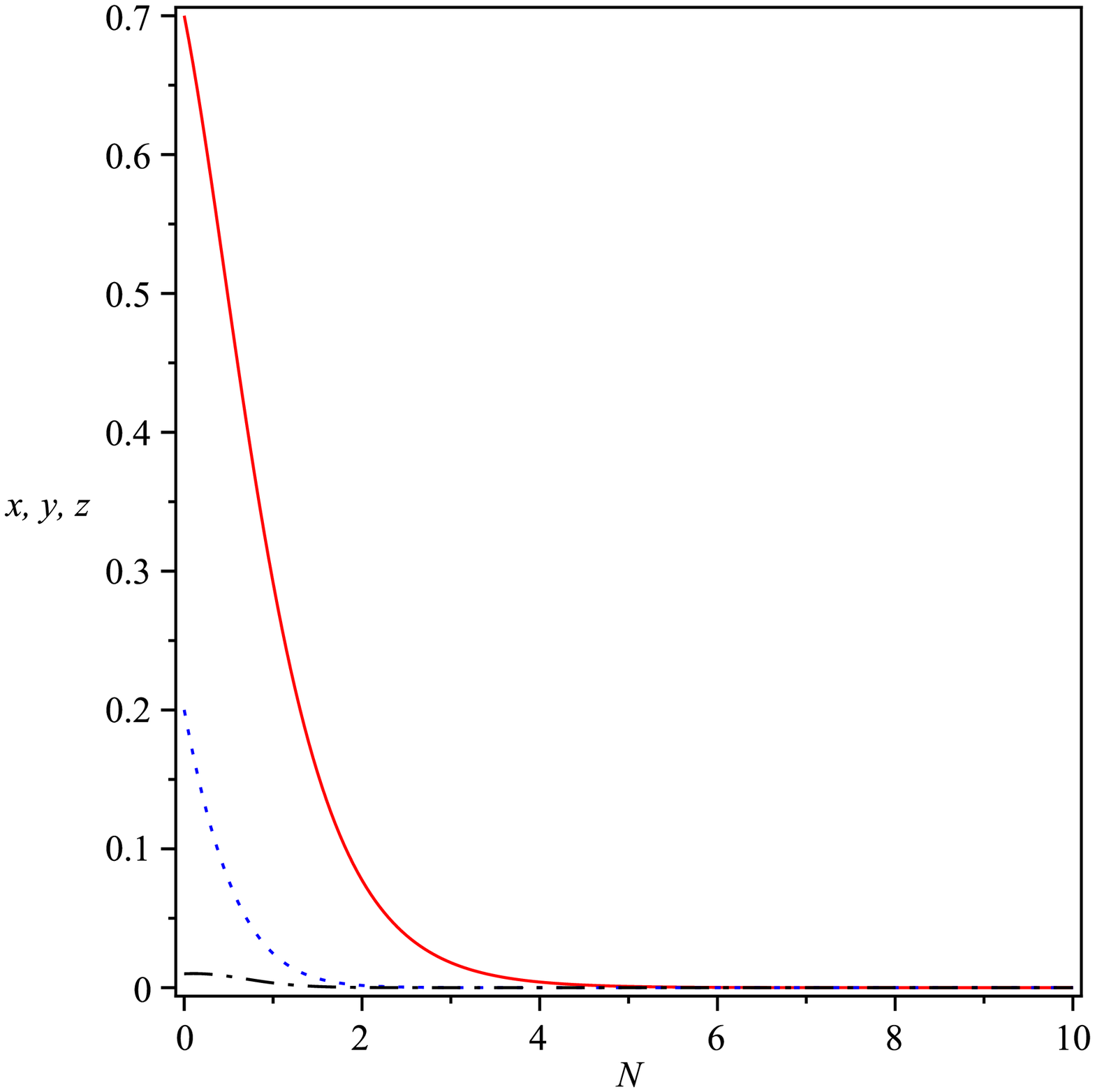}% scale goes from 0 to 1.
  \caption{Model III: variation of $x,y,z$ as a function of the $N=\ln(a)$. The initial
  conditions chosen are $x(0)=0.7,y(0)=0.3,z(0)=0.01$,
  $w_d=-0.5$ and $b=0.5$. }
  \label{16.eps}
\end{figure}

\begin{figure}
\centering
 \includegraphics[scale=0.4] {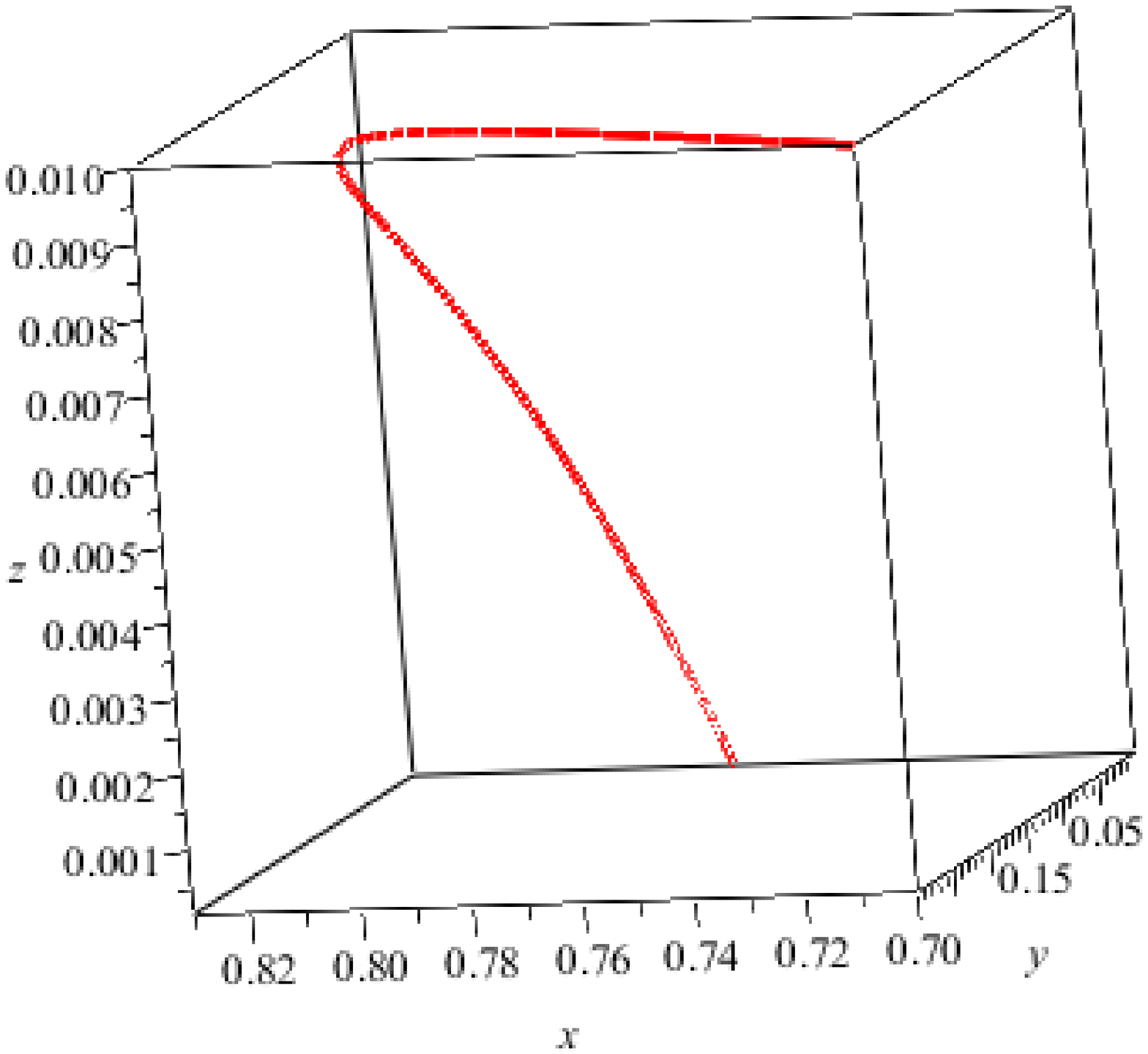}% scale goes from 0 to 1.
  \caption{ Model III: Phase space for $w_d=-1, b=0.5,\alpha=0.5$.}
  \label{17.eps}
\end{figure}

\begin{figure}
\centering
 \includegraphics[scale=0.3] {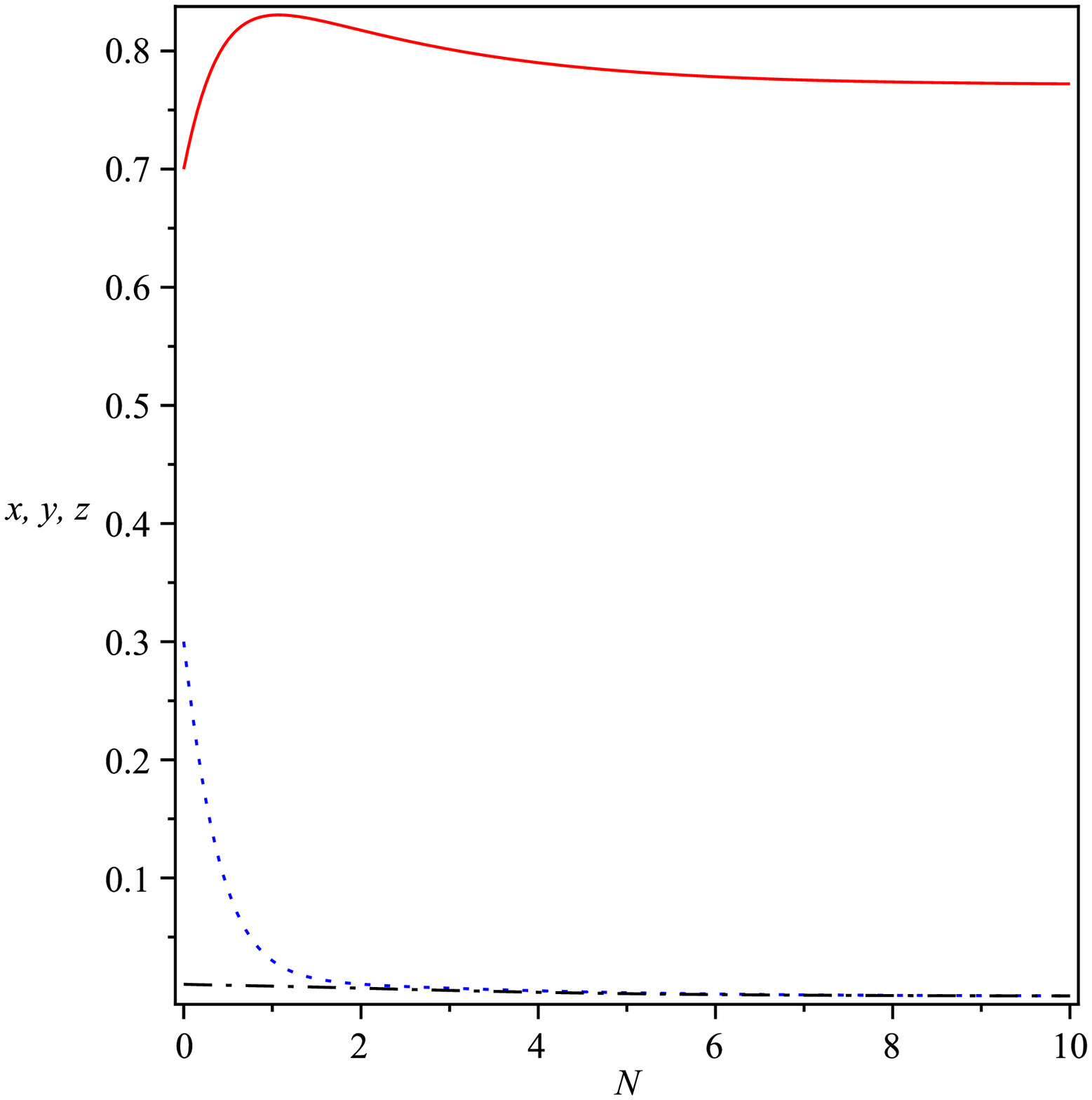}% scale goes from 0 to 1.
  \caption{Model III: variation of $x,y,z$ as a function of the $N=\ln(a)$. The initial
  conditions chosen are $x(0)=0.7,y(0)=0.3,z(0)=0.01$,
  $w_d=-1$ and $b=0.5$. }
  \label{18.eps}
\end{figure}

The eigenvalues of the Jacobian matrix for these critical points
are:

\begin{itemize}
\item Point $A_3:\lambda_1=-3, \lambda_2= -4,
\lambda_3= -3(1+w_d),$
\item Point $B_3:\lambda_1= 3, \lambda_2=-1,
\lambda_3=-3w_d$,
\item Point $C_3:\lambda_1=3w_d, \lambda_2=3(1+w_d), \lambda_3= 9\,b\alpha+9\,b-1+3\,w_{{d}}$,
\item Point $D_3:\lambda_1=4 ,  \lambda_2=1-9b-9b\alpha,\lambda_3=-9b\alpha-9b+1-3w_d.$
\end{itemize}
 $A_3$ is stable for $w_d>-1$. $B_3$ , $D_3$ are unstable.  $C_3$ is conditionally stable when $w_d<-1$, $b<\frac{1-3w_d}{9(1+\alpha)}$.

In figures (13-18), we have plotted the cosmological parameters of
model-III. In figures 13, 15 and 17, we show the attractor solutions
of the differential equations equation. In figure 14, we observe the
oscillatory behavior of dark energy and other cosmic components. It
shows that when DE energy density decays then corresponding
densities of dark matter and radiation increases and vice versa.
Figure 16 shows that all forms of energy densities vanish by $N\sim
4.5$ From figure 18, the radiation density stays zero while matter
energy density decreases and vanish by $N\sim2$. The dark energy
density increases by $N\sim2$ while it decreases and stays constant
at later epochs.

\subsection{Interacting Model - IV}

Consider another model with the interaction terms \cite{jamil8}
\begin{eqnarray}\label{11}
\Gamma_1&=&-3b\kappa^2 H^{-1}\rho_d\rho_r,\nonumber\\
\Gamma_2&=&3b\kappa^2 H^{-1}(\rho_d\rho_r-\rho_m\rho_r),\nonumber\\
\Gamma_3&=&3b\kappa^2 H^{-1}\rho_m\rho_r.
\end{eqnarray}
The system in (\ref{sys}) takes the form
\begin{eqnarray}
\frac{dx}{dN}&=&3x\Big(\frac{x+y+z+w_dx+w_r z}{1+\alpha}\Big)-3x-3w_dx-9bxz,\nonumber\\
\frac{dy}{dN}&=&3y\Big(\frac{x+y+z+w_dx+w_r z}{1+\alpha}\Big)-3y+9b(xz-yz),\\
\frac{dz}{dN}&=&3z\Big(\frac{x+y+z+w_dx+w_r
z}{1+\alpha}\Big)-3z-3w_rz+9byz.\nonumber
\end{eqnarray}
There are seven critical points:
\begin{itemize}
\item Point $A_4:\ (0,0,0)$,
\item Point $B_4:\ (1+\alpha,0,0)$,
\item Point $C_4:\ (0,1+\alpha,0) $,
\item Point $D_4: \ (0,0,1+\alpha)$,
\item Point  $E_4: \ \Big(\frac{1}{3}\frac{\frac{4}{3}w_d}{b(1+w_d)},\frac{4}{9b},-\frac{1}{3}\frac{1+w_d}{b}\Big)$,
\item Point  $F_4: \Big( \,{\frac {w_{{d}}+3\,b+3\,b\alpha-1/3}{3b}}   ,-\,{\frac { \left( -\frac{1}{3}+w_{{d}} \right)
 \left(w_{{d}}+3\,b+3
\,b\alpha-\frac{1}{3} \right) }{3b \left(
3\,b+3\,b\alpha-\frac{1}{3} \right) } } ,\,\\\nonumber{\frac
{w_{{d}} \left( -\frac{1}{3}+w_{{d}} \right) }{3b \left( 3\,b+3
\,b\alpha-\frac{1}{3} \right) }}\Big) $,
\item Point  $G_4: \ (0,\frac{4}{9b},-\frac{1}{3b})$.
\end{itemize}
 For points $A_4,B_4,C_4,D_4,G_4$, the eigenvalues of the Jacobian
 matrix are
\begin{itemize}
\item Point $A_4:\ \lambda_1=-3,\lambda_2=-4, \lambda_3=-3(1+w_d)$,
\item Point $B_4:\ \lambda_1=3w_d,\lambda_2=3(1+w_d),\lambda_3=3w_d-1$,
\item Point $C_4:\ \lambda_1= 3,\lambda_2=-3w_d,
\lambda_3=9b(1+\alpha)-1$,
\item Point $D_4:\ \lambda_1=3(1+w_d), \lambda_2=-9b-9b\alpha+1,\lambda_3=-3w_d-9b-9b\alpha+1$,
\item Point $G_4:\ \lambda_1=-3w_d, \\ \lambda_2=\frac {\sqrt {3}\sqrt {b \left( 1+\alpha \right)
\left( 3\,b\alpha+3\,b/3\alpha-4/9+3\,b+\,b \right) }}{b \left( 1+\alpha \right) },\\ \lambda_3=-{\frac {\sqrt {3}\sqrt {b \left( 1+\alpha \right)\left( 3\,b\alpha+3 \,b/3\alpha-4/9+3\,b+\,b \right) }}{b \left( 1+\alpha \right) }}$
\end{itemize}

It is observed that in this model, $A_4$ is stable for $w_d>-1$ , $B_4$ is stable for $w_d<-1$.  $C_4$ is unstable.  $D_4$ is stable for $w_d<-1,b>\frac{1}{9(1+\alpha)}$. It's not possible to determine the stability of the point $E_4$. But  $G_4$ is unstable.

\begin{figure}
\centering
 \includegraphics[scale=0.4] {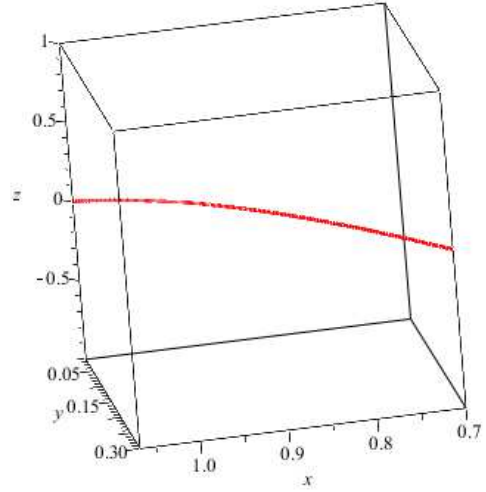}% scale goes from 0 to 1.
  \caption{Model IV: Phase space for $w_d=-1.2, b=0.5,\alpha=0.5$. }
  \label{19.eps}
\end{figure}

\begin{figure}
\centering
 \includegraphics[scale=0.3] {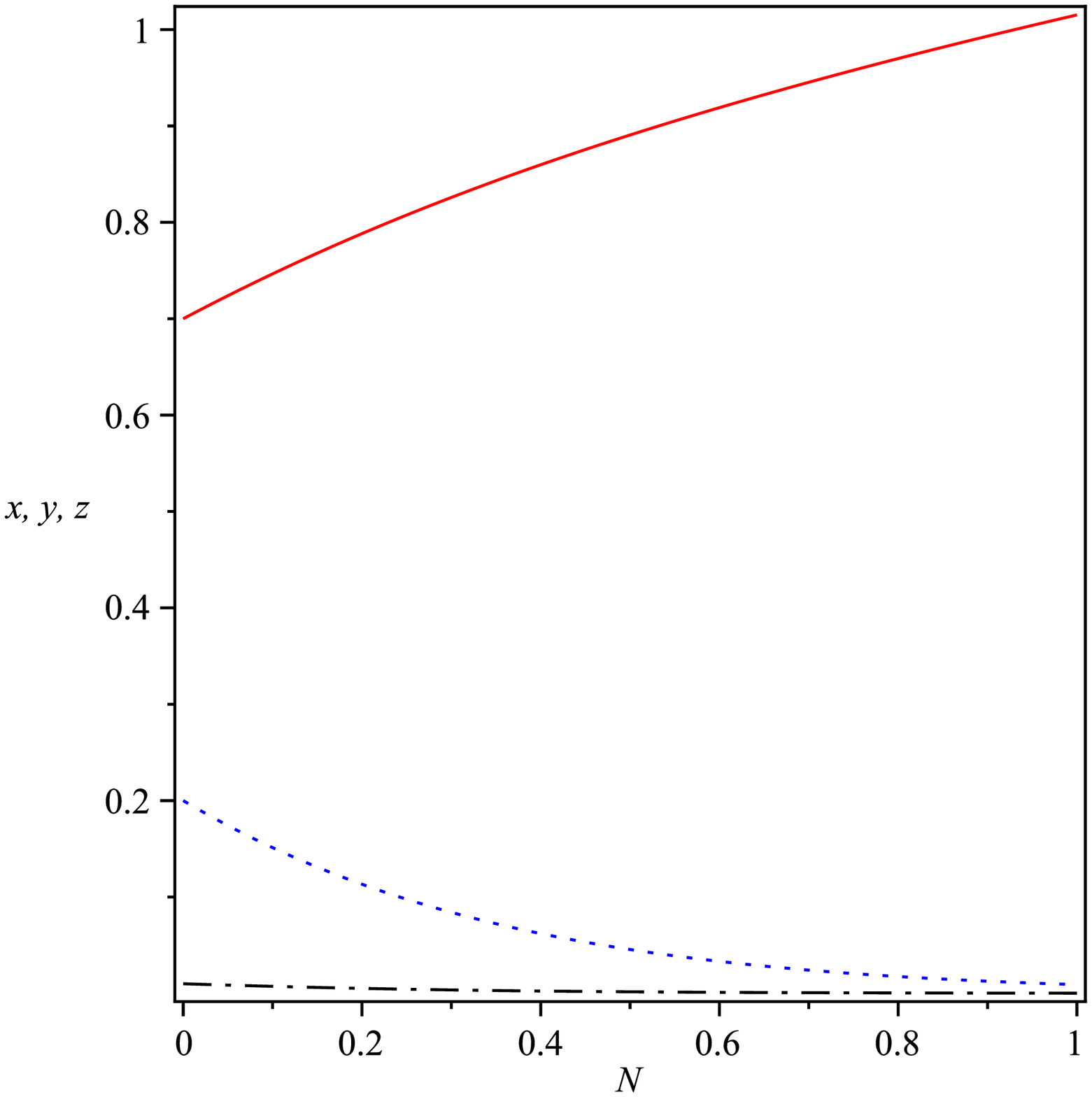}% scale goes from 0 to 1.
  \caption{ Model IV: variation of $x,y,z$ as a function of the $N=\ln(a)$. The initial
  conditions chosen are $x(0)=0.7,y(0)=0.3,z(0)=0.01$,
  $w_d=-1.2$ and $b=0.5$.}
  \label{20.eps}
\end{figure}

\begin{figure}
\centering
 \includegraphics[scale=0.4] {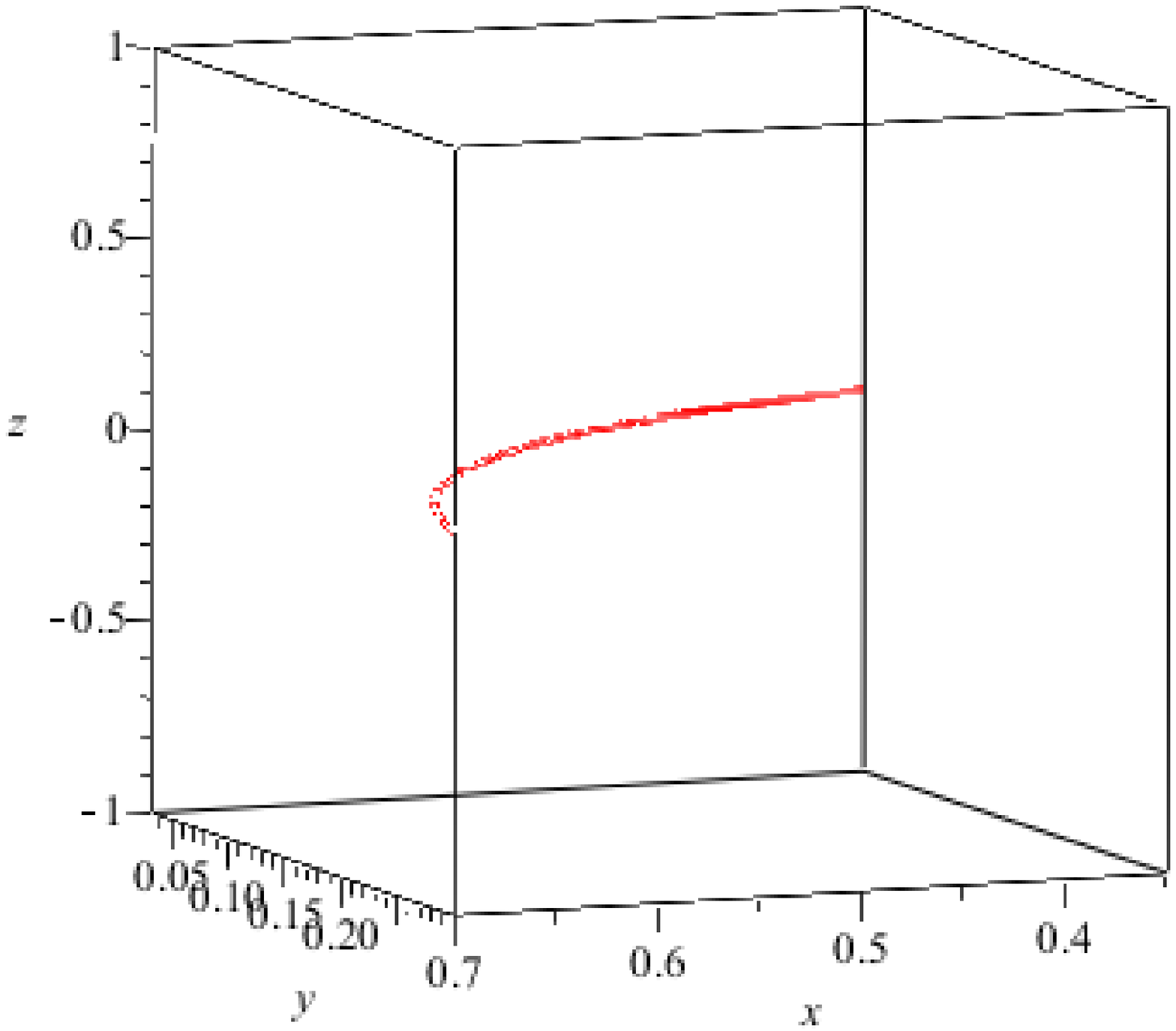}% scale goes from 0 to 1.
  \caption{ Model IV: Phase space for $w_d=-0.5, b=0.5,\alpha=0.5$.}
  \label{21.eps}
\end{figure}

\begin{figure}
\centering
 \includegraphics[scale=0.3] {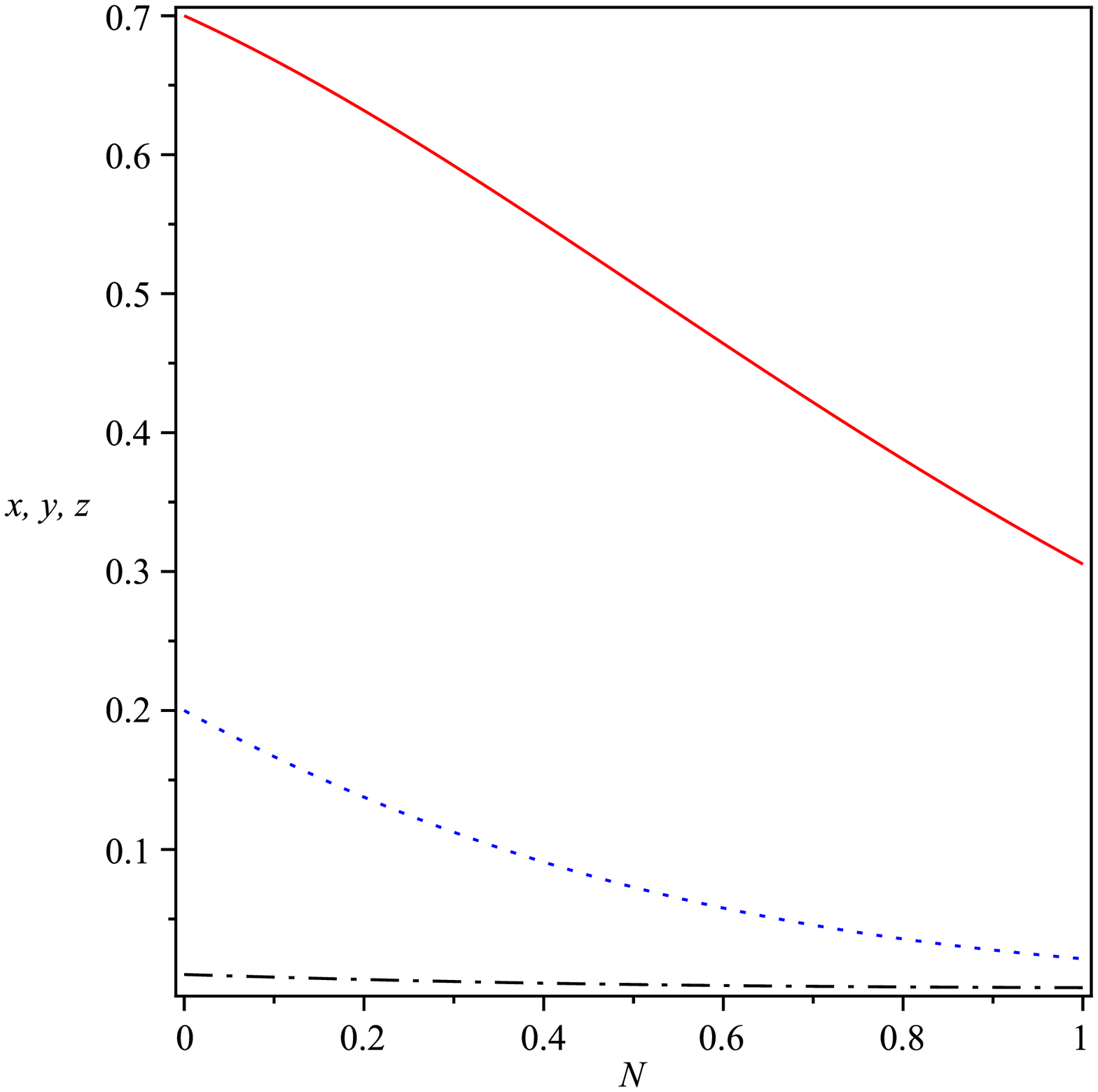}% scale goes from 0 to 1.
  \caption{Model IV: variation of $x,y,z$ as a function of the $N=\ln(a)$. The initial
  conditions chosen are $x(0)=0.7,y(0)=0.3,z(0)=0.01$,
  $w_d=-0.5$ and $b=0.5$. }
  \label{22.eps}
\end{figure}

\begin{figure}
\centering
 \includegraphics[scale=0.4] {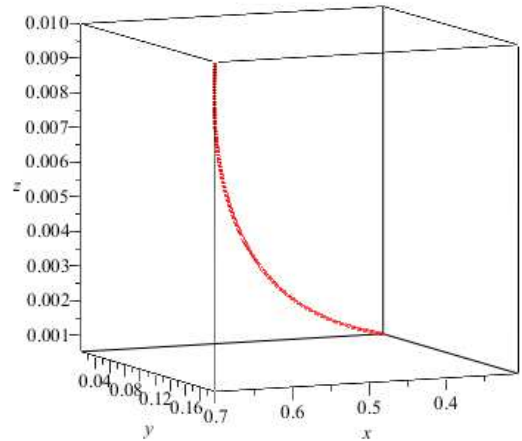}% scale goes from 0 to 1.
  \caption{ Model IV: Phase space for $w_d=-1, b=0.5,\alpha=0.5$.}
  \label{23.eps}
\end{figure}

In figures (19-23), we give description about model-IV. Figures
19,21,23 represent the phase space diagrams for different forms of
dark energy. Figures 22 and 23 shows that energy density of dark
energy decreases while in figure 20, the density of DE increases. It's the typical behavior of the phantom fields. Thus, our model predicts the correct evolutionary schem for the DE density in regime of phantom.

%\section{Lorentz oscillator in F(T) gravity}
%Let us we consider the following interaction terms
%\begin{eqnarray}\label{eq}
%\Gamma_1&=&-3\kappa^{-2}H^3[-3x(1+w_d)+3x\Big(\frac{x+y+z+w_d x+w_r z}{1+f_T+2T f_{TT}}\Big)-\sigma(y-x)],\nonumber\\
%\Gamma_2&=&-3\kappa^{-2}H^3[-3y+3y\Big(\frac{x+y+z+w_d x+w_r z}{1+f_T+2T f_{TT}}\Big)-x(\delta-z)+y]\label{23}\\
%\Gamma_3&=&-3\kappa^{-2}H^3[-3z(1+w_r)+3z\Big(\frac{x+y+z+w_d x+w_r z}{1+f_T+2T
%f_{TT}}\Big)-xy+\beta z],\nonumber
%\end{eqnarray}
%Substituting these expressions into (6) we get
%\begin{eqnarray}\label{eq}
%\frac{dx}{dN}&=&\sigma(y-x),\nonumber\\
%\frac{dy}{dN}&=&x(\delta-z)-y,\\
%\frac{dz}{dN}&=&xy-\beta z.\nonumber
%\end{eqnarray}
%It is famous Lorentz oscillator equation.

\section{Conclusion}
$f(T)$ gravity is a powerful and novel theory for explanation of the
acceleration expansion of the universe. In this paper we discussed
the stability of the interactive models of the dark energy, matter
and radiation in a FRW model, for a general $f(T)$ theory. We
derived the equations and show that why we have some attractor
solutions for some specific forms of the interactions. We
numerically integrate the equations and show that the evolution of
the dark energy density mimics three diffract behaviors phantom,
quintessence and cosmological constant in some interactive forms. We
like to comment that this interaction is purely phenomenological
required to meet some observational consequences. Since the phase
space of the system of the evolutionary equations has a definitive
end point, trackers are exist and depending on the initial
conditions, there are different kinds of the trackers.

\subsection*{Acknowledgment}
We would like to thank anonymous referees for giving useful comments to improve this paper.
M. Jamil and D. Momeni would like to thank the warm hospitality of Eurasian National University where this work was started and completed.

\end{document}